\documentstyle[editedbook,epsfig,epsf,graphicx]{mq}
\def\eg{{\it e.g.} }
\def\etal{{\em et al.} }
\def\ie{{\em i.e.} }

\def\cm2{cm$^2$ }
\def\se1{s$^{-1}$ }

\def\arcsec{\hbox{$^{\prime\prime}$} }
\def\degree{$^{\circ}$ } 

\begin{opening}
\title{SIMBOL--X, a new generation X--ray telescope for the 
0.5--70~keV range}
\author{Philippe Ferrando}
\institute{CE Saclay, DSM/DAPNIA/Service d'Astrophysique, F91191 
Gif-sur-Yvette France.}
\end{opening}

\runningtitle{SIMBOL--X}
\runningauthor{P.\ Ferrando}

\begin{document}
\vspace{-0.5cm}
\begin{abstract}
{\small SIMBOL--X is a high energy ``mini" satellite class mission 
that is proposed by a French-Italian-English collaboration for a launch 
in 2009. SIMBOL--X is making use of a classical X--ray mirror, of 
$\sim$~600~\cm2 maximum effective area, with a 30~m focal length in 
order to cover energies up to several tens of keV. This focal length 
will be achieved through the use of two spacecrafts in a formation 
flying configuration. This will give to SIMBOL--X unprecedented 
spatial resolution (20\arcsec HEW) and sensitivity in the hard X--ray 
range. By its coverage, from 0.5 to 70~keV, and sensitivity, SIMBOL--X 
will be an excellent instrument for the study of high energy processes 
in a large number of sources, as in particular accreting black-holes, 
extragalactic jets and AGNs.}
\end{abstract}

\section{Scientific case}
The study of the non thermal component in high energy astrophysics 
sources is presently hampered by the large gap in spatial resolution 
and sensitivity between the X--ray and $\gamma$--ray domains. Below 
$\sim$~10~keV, astrophysics missions like XMM--Newton and Chandra are 
using X--ray mirrors based on grazing incidence reflection properties. 
This allows to have an extremely good spatial resolution, down to 
0.5\arcsec for Chandra, and a good signal to noise thanks to the 
focussing of the X--rays onto a small detector surface. This 
technique has however an energy limitation at $\sim$~10 keV due to 
the maximum focal length that can fit in a single spacecraft. Hard 
X--ray and $\gamma$--ray imaging instruments, such as those on 
INTEGRAL to be launched in Oct.~2002 are thus using a 
different technique, that of coded masks. This non focussing 
technique has intrinsically a much lower signal to noise ratio than 
that of a focussing instrument, and does not allow to reach spatial 
resolutions better than $\sim$~10 arc minutes. This results for example 
in roughly 2 orders of magnitude of difference in point source sensitivity 
between X--ray and $\gamma$--ray telescopes.

This transition of techniques unfortunately happens roughly at the 
energy above which the identification of a non thermal component is 
unambiguous with respect to thermal emission. This obviously strongly 
limits the interpretation 
of the high quality X--ray measurements, and particularly that related 
to the acceleration of particles. Considered from the high energy side, this 
renders impossible the identification of the $\gamma$--ray emitters 
counterparts. As a single example of this fact, in relation with 
this workshop, one can cite the case of the SS~433/W50 system. The 
Eastern lobe is known to emit above 10~keV thanks to RXTE 
observations (Safi-Harb \& Petre, 1999), but the interpretation of 
this emission is strongly dependent on the assumptions made on the 
size of the emitting region which is completely unknown because of 
the very poor angular resolution of RXTE ($\sim$~1\degree, similar 
to the lobe size).

The SIMBOL--X mission is basically designed to extend the X--ray 
focussing technique to much higher energies, up to $\sim$~70 keV, \ie 
well beyond the transition between thermal and non thermal emissions. 
Offering a constant spatial resolution and a ``soft X--ray type"
sensitivity over the full energy range from 0.5 to 70~keV, SIMBOL--X 
will be an excellent instrument to elucidate the origin of the non thermal 
emission in accretion / acceleration astrophysical sites, both compact 
and extended.

\section{Mission concept}
\begin{figure}[hb]
\begin{minipage}{0.49\linewidth}
\centering
\epsfig{file=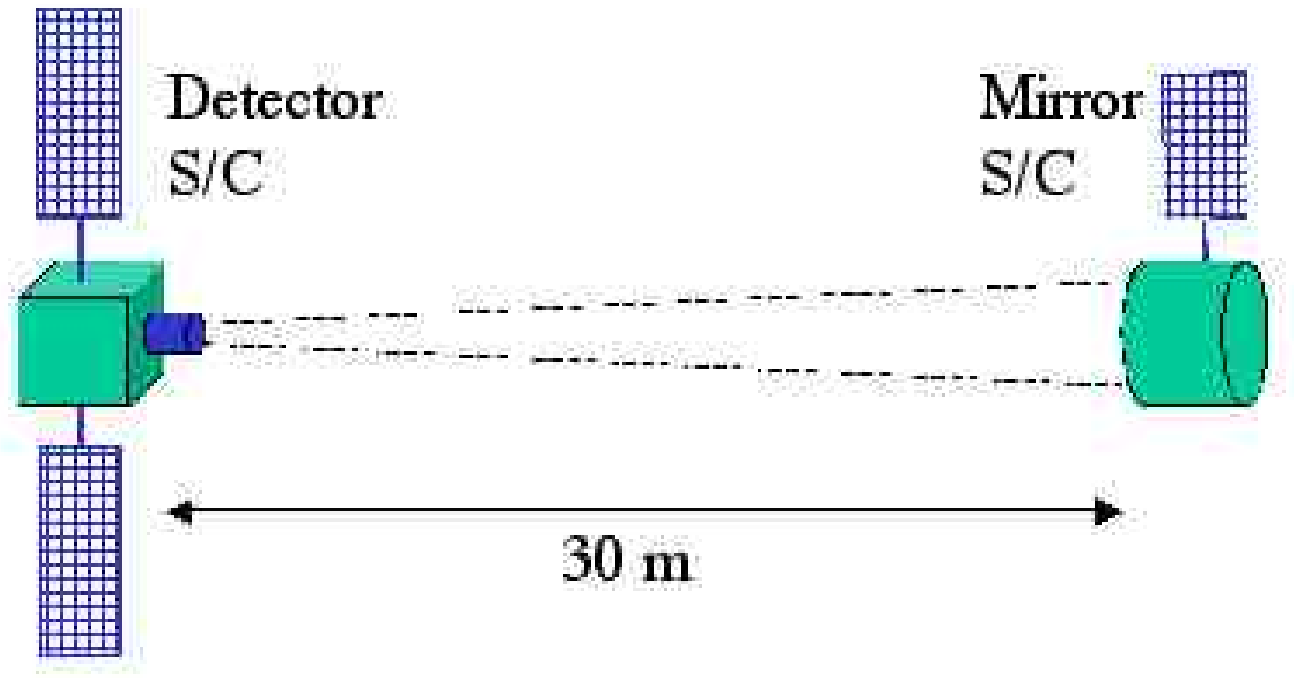,width=\linewidth}
\caption{SIMBOL--X two spacecrafts configuration.}
\label{fig:fly}
\end{minipage}
\hfill
\begin{minipage}{0.49\linewidth}
\centering
\epsfig{file=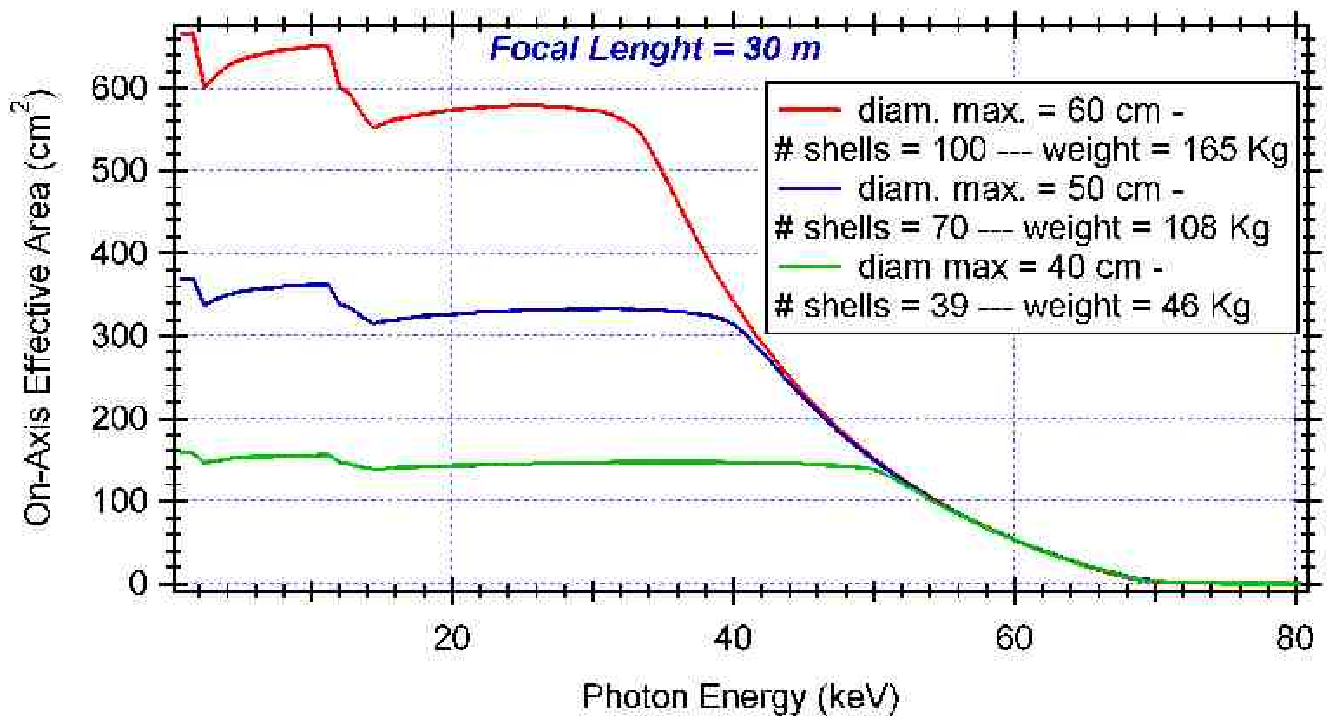, width=\linewidth}
\caption{Mirror effective area for different configurations. The 
baseline for SIMBOL--X is the top curve.}
\label{fig:mirror}
\end{minipage}
\end{figure}

SIMBOL--X is built on the classical design of a Wolter~I optics 
focussing X--rays onto a focal plane detector system. 
The gain in maximum energy is achieved by having a long 
focal length, of 30 metres \ie 4 times that of XMM--Newton 
mirrors. Since this cannot fit in a single spacecraft, the mirror 
and the detectors will be flown on two separate spacecrafts, in a 
formation flying configuration, as sketched on Fig.~\ref{fig:fly}. We 
shortly detail below each part of this system.

\vspace{-0.5cm}
\subsection{Mirror}
\label{sec:mirror}
The focussing optics will be a nested shells Wolter~I configuration mirror. 
Building on the experience acquired on Beppo--SAX, Jet--X, SWIFT, 
ABRIXAS and XMM--Newton mirrors, it will be made following the
Nickel electroforming replication method (\eg Citterio \etal 2001).
The current design is to have a 108 shells mirror, with an outer 
diameter of 70~cm (like one XMM--Newton module), and an angular 
resolution of 20 arcseconds of Half-Energy-Width. The coating will be 
Pt, in order to increase the high energy response w.r.t.\ a more 
classical Au coating. The focal length will be 30 metres.

Figure \ref{fig:mirror} shows the effective area as a function of 
energy, for the baseline design. It has roughly a constant value of 
600~cm$^2$ up to about 35~keV, before starting to decrease and fall 
below one cm$^2$ above more than 70~keV. The field of view (FOV) 
will be of 6 arcminutes at 50~\% vignetting.

\subsection{Focal plane detectors}
\label{sec:detector}
The focal plane detector system has obviously to be a spectro-imager 
covering the full FOV and sensitive to the full energy range provided 
by the mirror, with a ``reasonable" spectral resolution below 10~keV 
for measuring lines, particularly that of Iron. Given the above mirror 
characteristics, the focal plane detector must be of 6~cm diameter 
with 500~$\mu$m of maximum pixel size (to provide an oversampling of 
the Point Spread Function).

The baseline focal plane responding to these constraints is made of 
an array of CCD detectors directly on top of a CdZnTe pixellated 
detectors array. This is completed by an optical blocking filter, 
and an active anticoincidence shield. Photons of energy less than 
$\sim$~15~keV will stop in the CCDs, whereas higher energy photons 
will go through the CCDs to be detected in the CdZnTe array. CCDs 
with 250~$\mu$m depletion depth and 2~mm thick CdZnTe, currently under 
tests in Leicester and CEA/SAp resp., are perfectly fitting the 
SIMBOL--X requirements.

\subsection{Formation flying constraints and orbit}
Both mirror and detector spacecrafts will be of the ``mini-satellite" 
class (500 kg max). To keep a constant image quality requires the 
following constraints on the spacecraft relative positionning : i) their
distance must be kept constant within 1~cm, ii) their positionning 
perpendicular to the optical axis must be kept within 1~cm, and monitored 
with a 0.5~mm accuracy, and iii) the angular stability must be better than 
1~arcmin, and monitored to 3~arcsec. In order to minimize the 
differential forces between the 2 spacecrafts, as well as to allow 
uninterrupted observations of variable sources, SIMBOL--X will 
be put in orbit around L2.

\section{Sensitivity}
\begin{figure}[ht]
\epsfig{file=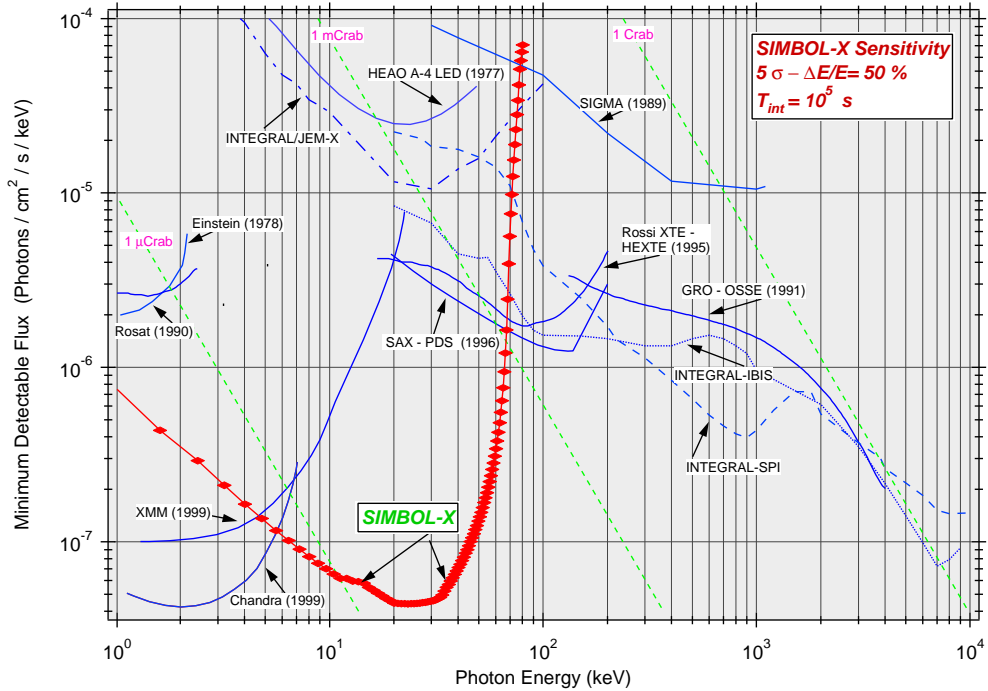,width=\linewidth}
\caption{SIMBOL--X sensitivity to point sources, compared to past and
present X and $\gamma$--ray telescopes.}
\label{fig:sensi}
\end{figure}

In order to calculate the signal to background ratio for astrophysical 
sources, we have used the radiation background spectrum calculation of 
Ramsey (2001) for the HXT CdZnTe detector, with veto, envisionned for 
Constellation~X (also at the L2 orbit), a configuration very similar to that 
of SIMBOL--X. We have also modelled the astrophysical diffuse background. 
With these, we have derived the sensitivity of SIMBOL--X to point 
sources, or equivalently to diffuse emission in a 1 arcmin 
diameter region. This is presented in Fig.~\ref{fig:sensi} with respect to 
other instruments. As expected for an X--ray focussing telescope, 
the SIMBOL--X sensitivity curve has roughly the shape of the 
XMM--Newton and Chandra curves (which have no diffuse component 
here), but is displaced by about a decade in energy. 
SIMBOL--X is about $\sim$~100 times better than existing instruments 
in the 10 to 35~keV range, and has a sensitivity equivalent to 
INTEGRAL/IBIS at $\sim$ 70 keV.

We have also used this background model in order to simulate a number 
of observations that cannot be detailed here. We simply mention two 
examples to illustrate the SIMBOL--X sensitivity. On the supernovae 
remnant side, a detailed map of Cas~A above 20~keV can be done in 
100~ksec, and the spectrum of its brightest 1~arcmin$^2$ part is 
significant up to 50~keV. On the AGN jet side, the spectrum of the 
Pictor~A hot spot at 4~arcmin from the nucleus (well isolated with 
SIMBOL--X optics) can be significantly measured up to 40~keV in 
50~ks of observation.

\section{Collaboration and schedule}
The SIMBOL--X mission is a collaboration involving France (CEA/Saclay, 
PI Institute, and the Grenoble and Meudon 
observatories), Italy with Brera Observatory, and United Kingdom with 
Leicester University. At this early stage, the participation of each country
which has still to be secured is envisionned to be the following. 
France will build the detector spacecraft, the high energy part of the
focal plane detector, and will take care of the formation flying aspects. Italy 
is in charge of building the mirror and the mirror spacecraft. United 
Kingdom is in charge of the CCD part of the focal plane.

Finally, as SIMBOL--X does not involve new difficult technological 
development neither on the detector side nor on the mirror side, and as
the formation flying constraints are rather well within current 
investigations in this domain, a relatively short development time can 
be envisionned. Launching SIMBOL--X before the end of the decade 
would provide an excellent scientific preparation to the much larger
observatories Constellation~X and XEUS that are scheduled later. 

SIMBOL--X has been presented to the Astrophysics group of CNES, the 
french space agency, with a proposed launch date of early 2009. It has 
been selected in June 2002 for beginning a phase A study. 


\end{document}